\documentclass[twocolumn,showpacs,aps,prl]{revtex4}
\usepackage{amssymb}
\usepackage{graphicx}
\usepackage{dcolumn}
\usepackage{bm}
\usepackage{color}

\begin{document}

\title{Heat to Electricity Conversion by a Graphene Stripe with Heavy Chiral Fermions}
\author{S.~E.~Shafranjuk}
\affiliation{Department of Physics and Astronomy, Northwestern University, Evanston, IL 60208}

\date{\today }


\begin{abstract}
A conversion of thermal energy into electricity is considered in the electrically polarized graphene stripes with zigzag edges where the heavy chiral fermion (HCF) states are formed. The stripes are characterized by a high electric conductance $G_e$ and by a significant Seebeck coefficient ${\cal S}$. The electric current in the stripes is induced due to a non-equilibrium thermal injection of  ``hot"  electrons. This thermoelectric generation process  might be utilized for building of thermoelectric generators with an exceptionally high figure of merit $Z \delta T \simeq 100 >> 1$ and with an appreciable electric power densities $\sim $ 1~MW/cm$^2$.
\end{abstract}

\maketitle

The heat energy can be immediately converted into electricity by   thermoelectric generators (TEG).~\cite{TEG-book,Dressel,P-Kim,Shafr-TEG} 
However, available solid state TEGs typically have a low figure of merit $Z \delta T << 1$. In this paper we suggest an approach which allows improving the TEG characteristics considerably. The key element of the our thermoelectric generator (TEG) is the graphene stripe $\mathcal{G}$ with atomic zigzag edges~\cite{Fertig} polarized by the transverse electric field $\mathbf{E}$ (see Fig.~\ref{fig2}). The graphene TEG utilizes the gate voltage-controlled heavy chiral fermion (HCF) states formed inside $\mathcal{G}$ (see Section~II) and it also benefits from specially introduced electric/heat current filtering. 
For the sake of convenience we use the figure of merit defined as a ratio $Z\delta T = Q_{\rm el}/Q_{\rm th}$ where $Q_{\rm el}$ is the generated electric power, $Q_{\rm th} $ is the heat flow energy carried by  the lattice oscillations, phonons; $\delta T$ is the temperature difference between the ``hot" and ``cold" ends.~\cite{P-Kim,Shafr-TEG,Pop,Cahill}
The design strategy outlined in Section III is illustrated by an expression for the figure of merit $Z \delta T=G_e{\cal S} ^{2} \delta T/\Lambda $ for a symmetric TEG.~\cite{TEG-book,Dressel} Here ${\cal S} $ is the Seebeck coefficient, $G_e $ is the electric conductance, 
$\Lambda =\Lambda _{\mathrm{e}}+\Lambda _{\mathrm{ph}}$ is the thermal conductance due to the electrons (e) and phonons (ph), $\delta T$ is the temperature difference between the ``hot'' and ``cold" sides. 

In typical TEG devices,~\cite{Cahill} $\Lambda _{\mathrm{ph}}/\Lambda _{\mathrm{e}}\sim 10^{3}-10^{4}$, providing that $\Lambda \simeq \Lambda _{\mathrm{ph}}>>\Lambda _{\mathrm{e}}$. Since normally $\Lambda >>$ $G_e {\cal S} ^{2}\delta T$, one gets $Z \delta T{<<}1$. Initial thermoelectric measurements of carbon nanotubes (CNT) also have not revealed their potential as efficient TEGs. The reasons why CNT do not show any spectacular $ZT$ are as follows. ({\it a})~The heat conductance $\Lambda_{\rm ph}$ due to phonon transport is too high in CNT and graphene. ({\it b})~The electronic transport properties were not appropriately optimized to achieve a plausible $Z \delta T$. In the TEG reported in Sections~II,III, instead of regular electrons and holes, we exploit heavy chiral fermions (HCF) to boost the electric conductance $G_e \propto \sqrt{m^*/m_e}$ and Seebeck coefficient ${\cal S} $. The HCF states produce incredibly sharp spectral singularities in the electron density of states (DOS). In contrast to a pristine graphene, the effective electron mass $m^*$ in vicinity of HCF peaks in $\mathcal{G}$-TEG exceeds the free electron mass $m_e$ by a huge factor $m^*/ m_e \sim 10^2-10^4 >> 1$ which allows increasing the $Q_{\rm el}$ considerably. In Section~IV, A-I, A-II we also utilize special multilayered metallic electrodes with an aim of decimating $\Lambda _{\mathrm{ph}}$. It allows reducing the phonon part $Q_{\rm th}$ of heat energy flow by factor $\sim 10^{-2}-10^{-3}$. The combined approach discussed in Sections~V,VI yields $Z\delta T  = Q_{\rm el}/Q_{\rm th} {\geq }1$, or even $Z \delta T >> 1$, which manifests a significant improvement over existing TEG devices.

Our approach exploits a non-equilibrium thermal injection of electrons and holes from a ``hot" metallic electrode into HCF energy levels formed inside the narrow (10 nm - 100 nm) graphene stripe. Sharp distinctive HCF levels at energies $E_{p}=\pm \Delta $ result from
reflections at the atomic zigzag edges~\cite{Fertig} of the graphene stripe 
$\mathcal{G}$. The $\mathcal{G}$-stripe is polarized in the transverse $%
\hat{y}$-direction by a finite electric field $\mathbf{E}\neq 0$ which
controls the value $\Delta $. The origin of the HCF
resonances is illustrated in terms of a mere mean field approach 
\cite{Shafr-TEG,Ando}. The method~\cite{Shafr-TEG,Ando} describes the
electronic excitation spectrum $E_{p}$ of the graphene stripe in terms of
the Dirac equation for bipartite sublattices. One adds a symmetry breaking
term $\propto \Delta $ into the Hamiltonian%
\begin{eqnarray}
\mathcal{H}=-i\hbar v\left( \left( \hat{\sigma}_{x}\otimes \hat{1}\right)
\partial _{x}+\left( \hat{\sigma}_{y}\otimes \hat{\tau}_{z}\right) \partial
_{y}\right) \nonumber \\
 + U\left( x\right) \left( \hat{1}\otimes \hat{1}\right) +\Delta
\left( \hat{\sigma}_{z}\otimes \hat{\tau}_{z}\right)   \label{H_Dirac}
\end{eqnarray}%
where $v=8.1\times 10^{5}$m/s$\simeq c/300$ is the massless fermion speed, $%
\hat{\sigma}_{i}$ and $\hat{\tau}_{k}$ are the Pauli matrices, $\otimes $ is
the Kronecker product, $\{i,k\}=1\dots 3$, and $\Delta = e \vert {\bf E} \vert W/2= e V_{\rm SG}/2= e (V_{\rm GA} - V_{\rm GB})/2 $, $V_{\rm SG}$ is the split gate voltage, $W$ is the stripe width. The pseudospin polarization~\cite{Soriano} means that the electric charge is depleted on one zigzag edge while is accumulated on the other. Then an electric dipole is formed as soon as $V_{\rm SG} = 2\Delta/e \neq 0$. At $%
V_{\rm SG} =0$ there are two flat bands giving rise to a large DOS right
at the Fermi energy and being associated to zigzag edge states\cite{Fertig}.
When $V_{\rm SG} \neq 0$, the
resulting pseudo-spin polarization driven by the $\Delta $ term in Eq. (\ref%
{H_Dirac}) yields an excitation spectrum 
$ \varepsilon _{\pm }=\pm v\sqrt{%
k^{2}+q^{2}+\left( \Delta /v\right) ^{2}}\allowbreak \pm U$ characterized by
the energy gap\textbf{\ }$2\Delta $. 
For zigzag atomic edges, the electron wavefunction $\Psi $ satisfies the boundary conditions $\Psi _{\mathrm{A}}\left( 0\right) =\Psi _{\mathrm{A}^{\prime }}\left( 0\right) = 0$ at $y = 0$
and $\Psi _{\mathrm{B}}\left( W\right) = \Psi _{\mathrm{B}^{\prime }}\left(
0\right) = 0$ at $y = W$. It gives~\cite{Fertig} the
transversal quantization condition (QC) in the form 
\begin{equation}
k = q_{p}/\tan \left(
q_{p}W\right)
\label{trans}
\end{equation}
where 
\begin{equation}
q_{p} = \pm \sqrt{\left( E_{p}\mp U\right)
^{2}/v^{2}-k^{2}-\left( \Delta /v\right) ^{2}}.
\label{dispers}
\end{equation}
The QC actually determines
the localized energy levels when is being solved in respect to the
excitation energy $E_{p}$. Thus, for a finite $\Delta \neq 0$, the graphene
zigzag stripes $\mathcal{G}$ are band insulators with the pseudospin
polarization. The steady state energy spectrum of the $\mathcal{G}$-stripe features two flat bottom bands $E_{p}=\pm \Delta $ which represent the highest occupied and lowest un-occupied bands. The flat bands correspond to a very high effective electron mass $m^* = (10^2 - 10^5) m_e >> m_e$.

The sign of $U(x) $ determines whether the HCF charge carriers inside of each FET$_{\mathrm{L,R}}$ are electrons or holes. In Figs.~1b, the $V_{\mathrm{G}}\left( x\right)$ profile is step-wise, so $V_{\mathrm{G}}\left( x\right) = - U_0/\alpha_{\rm G}$ for $x < -L_{0}/2$ (inside FET$_{\mathrm{L}}$) and $V_{\mathrm{G}}\left( x\right) = U_0/\alpha_{\rm G} $ for $x > L_{0}/2$ 
(inside FET$_{\mathrm{R}}$). Here $U_0$ is the shift of the electron electrochemical potential and $\alpha_{\rm G} $ is the back gate efficiency. The $\mathcal{G}$-stripe RR section located immediately under the central hot H-electrode ($-L_{0}/2<x<L_{0}/2$) remains neutral, since $V_{\mathrm{G}}\left( x\right) = 0$. This part of the $\mathcal{G}$-stripe serves as an electron-hole recombination region (RR). 
\begin{figure}
\includegraphics[width=75 mm]{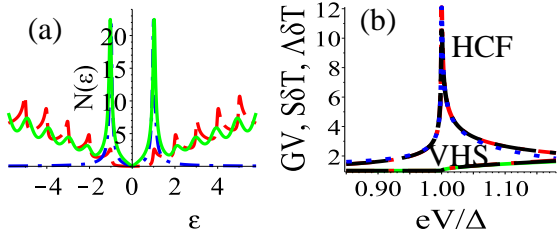}
\caption{{Color online. (a) The HCF singularity in the electron DOS $N(\varepsilon )$ at energy $\varepsilon = \Delta$ (solid green curve) as compared to the VHS singularities at  $\varepsilon = \Delta_{\rm VHS}^{(n)}$ (dash red curve) for $\gamma = 0.05 \Delta $. Blue dot-dash curve shows the fitting function $1/(\varepsilon - \Delta)^{1.3} $. (b)  The comparison plot showing an increase of thermoelectric current $G V$, the Seebeck voltage ${\cal S} \delta T$, and the electron heat current $\Lambda_{\rm e} \delta T$ when the source drain voltages $V_{\rm SD}$ matches either HCF  (correspondly shown by cyan, red, and black curves) or VHS  (shown by magenta, green, and blue curves) singularities.}} 
\label{fig1} 
\end{figure}

A $\mathcal{G}$-stripe with zigzag edges produces much stronger singularities (we call them HCF peaks) than the Van Hove singularities (VHS peaks) do. The VHS peaks in DOS have the form $\sim 1/\sqrt{\varepsilon - n \Delta_{\rm VHS}} $ where $n$ is an integer number. The VHS level spacing is $\Delta_{\rm VHS} \simeq h v_{\rm F}/\pi W$ where $W$ is the  $\mathcal{G}$-stripe width. The HCF  peaks are considerably sharper (in vicinity of the gap edge they are $\sim 1/(\varepsilon - \Delta)^\alpha $ where $\alpha = 1.3$) than the Van Hove peaks which are $\sim 1/\sqrt{\varepsilon - n \Delta_{\rm VHS}} $ and are relatively weak. The weak VHS singularity in DOS causes only very minor anomalies in the electric conductance and Seebeck coefficient as illustrated by Fig.~\ref{fig1}a. However, when the source drain bias voltage $V_{\rm SD}$ matches  $\Delta /e$, a much stronger HCF-singularity gives very sizable increase of the electric transport coefficients $G$, ${\cal S}$ and $\Lambda_e $ along the $\mathcal{G}$-stripe. That condition ensures a considerable thermoelectric effect taking place when $V_{\rm SD} = \Delta /e$. The DOS divergencies are smoothed by setting $\varepsilon \rightarrow \varepsilon + i \gamma$. However, if the source drain voltage  $V_{\rm SD}$ matches the weaker VHS, the change of DOS is rather modest. One see that from Fig.~\ref{fig1}a where the comparison between the HCF peaks (solid green curve) and VHS peaks (dash red curve) in DOS is presented. The HCF peak at $\varepsilon = \Delta $  is much sharper than the other peaks (which actually are the VHS singularities) on the same green curve at higher energies $\varepsilon > \Delta $  and $\varepsilon >> \Delta $. The 1$^{\rm st}$ HCF peak at $\varepsilon = \Delta $ is also sharper than the regular VHS peaks in red dash curve. In addition to the two curves with the HCF (solid green) and VHS (dash red) singularities we also plot a fitting curve $\propto 1/(\varepsilon - \Delta)^{1.3}$ (see blue dash-dot curve). Further comparison between contribution the two different kinds of singularities, HCF and VHS, is presented in Fig.~\ref{fig2}b. One might notice a considerable increase of thermoelectric current $I_{\rm th} = G V$ (cyan), the Seebeck voltage $V_{\rm th} = {\cal S} \delta T$ (red), and the electron heat current $Q_{\rm th} = \Lambda_{\rm e} \delta T$ (black curve) when the source drain voltage heats the HCF singularity ($V_{\rm SD} = \Delta/e$).  The HCF-induced icrease of $G V$, ${\cal S} \delta T$, and $\Lambda_{\rm e} \delta T$ is much stronger than a modest VHS-induced rise of the same characteristics (shown respectively with magenta, green, and blue curves) taking place at the source drain voltages $V_{\rm SD} = n \Delta_{\rm VHS}/e$.

By applying the split gate voltage $V_{\rm SG} \neq 0$ one might achieve the HCF level spacing as much as $\Delta = \alpha_{\rm SG} V_{\rm SG} = 10-100$~meV where $\alpha_{\rm SG} $ is the split gate efficiency factor.
The electron level broadening $\gamma =\left( 0.01-0.001\right) \Delta $ corresponds to $\gamma =0.2-1$~meV or $\gamma =2-10$~K. When using $\gamma =0.2$~meV, it corresponds to the electron mean free path $l_{\varepsilon }=\hbar
v_{F}/\gamma =3.3$~$\mu $m which is consistent with experiments with anealed
graphene samples. \cite{Bolotin} In principle, in ``clean" graphene samples one can get even the electron mean free path $l_{\varepsilon } = \hbar v_{F}/\gamma = 15 $~$\mu $m which corresponds to $\gamma =0.04$~meV. Then for $\Delta =100$~meV one obtains $\gamma /\Delta =0.0004$. The effective electron mass $m^{\ast }$ is evaluated in the tight binding model as $m^{\ast } \simeq \hbar^2 /(\gamma a^2)$ where $a$ is the graphene lattice constant. This gives $m^{\ast }/m_{e}=3\times 10^{5}$ which corresponds to the DOS enhancement factor $\sqrt{m^{\ast }/m_{e}}=2 \times 10^{2}$.

The VHS spacing in carbon nanotubes by diameter $\sim $2.5~nm is 0.4~eV which corresponds to temperatures $\sim $4600~K. That spacing much exceed the temperature broadening which for temperatures of interest $T \sim 300 - 600$~K typically is always much narrower than 4600~K. Therefore prevailing contribution at temperatures  $T <$ 600~K comes from one subband only. It might be different for wide graphene stripes with width $W >$~100~nm where the VHS spacing would be about $\sim $ 300~K. However the contribution of  HCF state into $Z T$ is always orders of magnitude stronger than the impact of  VHSs from higher energy bands.

Let us consider the TEG cell sketched in Fig.~\ref{fig2}a (we call it $\mathcal{G}$-TEG) and its diagrams shown in Fig.~\ref{fig2}b. 
The $\mathcal{G}$-TEG cell consists of two field effect transistors,~\cite{Shafr-Review,Lemme,Xu,Perebei,Kim}
left (FET$_{\mathrm{L}}$) and right (FET$_{\mathrm{R}}$), both being
fabricated on the same section of the graphene stripe $\mathcal{G}$. 
The two transistors, FET$_{\mathrm{L}}$ and 
FET$_{\mathrm{R}}$, are connected 
electrically in a sequence, but thermally in parallel (see Fig.~\ref{fig2}e).~\cite{Shafr-TEG} The edges of each $\mathcal{G}$-stripe have an atomic zigzag shape.~\cite{Fertig} The opposite edges of the same stripe are terminated by carbon atoms belonging to two distinct 
sublattices A and B.~\cite{Soriano} The split-gate voltage $V_{\rm SG} = V_{\rm GA} - V_{\rm GB} $ polarizes the $\mathcal{G}$-stripe in the transverse $y$-direction. The left (G$_{\mathrm{L}}$) and 
right (G$_{\mathrm{R}}$) back-gate electrodes are controlling the concentration and type of the charge carriers inside the left ($x < -L_{0}/2$) and right ($x > L_{0}/2$) $\mathcal{G}$-TEG sections, which are forming FET$_{\mathrm{L}}$ and FET$_{\mathrm{R}}$, correspondingly. The local gate voltage $V_{\mathrm{G}}\left( x\right) $ induces an $x$-dependent change $U(x)$ of the electrochemical potential along the $\mathcal{G}$-stripe as shown at the bottom of Fig.~\ref{fig2}b.

\begin{figure}
\includegraphics[width=85 mm]{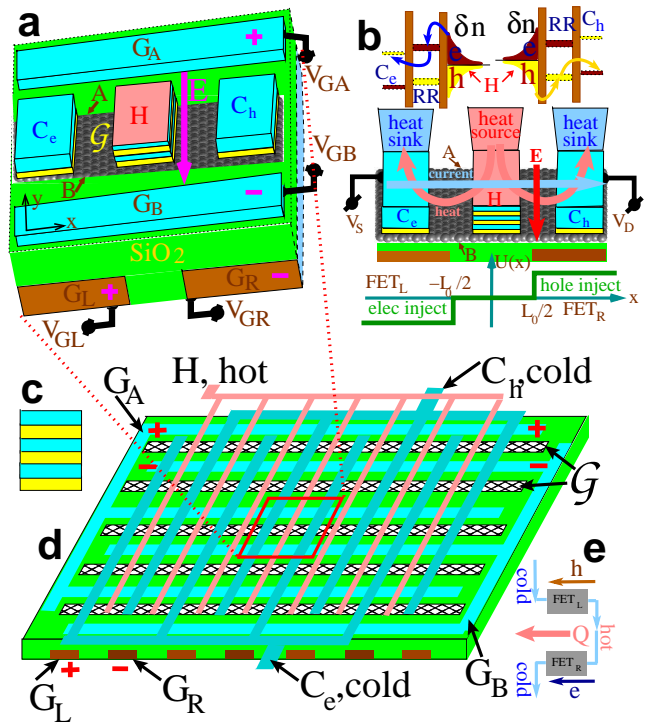}
\caption{{Color online. (a)~The $\mathcal{G}$-TEG cell which consists of a graphene  $\mathcal{G}$ stripe with zigzag edges deposited on a dielectric substrate. One edge of  $\mathcal{G}$ is terminated by atoms of A sublattice while the opposite edge by the B sublattice atoms. The $\mathcal{G}$-stripe is polarized in $\hat y$-direction by the transverse electric field $\vert {\bf E} \vert = V_{\rm SG}/W = (V_{\rm GA} - V_{\rm GB})/W$ ($W$ is the spacing between $G_{\rm A}$ and $G_{\rm B}$) which is created by the split gate electrodes G$_{\rm A}$ and G$_{\rm B}$. Two back gate electrodes G$_{\rm L}$ and G$_{\rm R}$ having opposite polarity induce a step-wise change of the electrochemical potential $U(x)$ versus coordinate $x$. (b)~The energy diagram (top panel), the heat and electric current diagram (central panel), and the $U(x)$ profile along the $\hat x$-direction inside $\mathcal{G}$ (bottom panel). (c)~The multi-layered H electrode where the phonon density of states $F_q (\omega )$ in adjacent layers is mismatched. (d)~The TEG device consists of the interdigitated split-gate combs G$_{\rm A(B)}$ on one hand and graphene comb $\mathcal{G}$ on the other hand. The H comb serves as the heat source while the G$_{\rm e}$ and G$_{\rm h}$ combs represent the heat sinks. The back gate electrode combs G$_{\rm L}$ and G$_{\rm R}$ provide the dynamic, coordinate-dependent $n$- and $p$-doping of  $\mathcal{G}$. (e)~The heat (${\bf Q}$) and electric (${\bf J}$) currents in the ${\cal G}$-TEG involving the field effect transistors (FET$_{\rm L,R}$).}} 
\label{fig2}
\end{figure}

The whole TEG circuit represents a 2D array formed by three interdigitated
combs $\mathcal{G}$, G$_{\mathrm{A}}$, and G$_{\mathrm{B}}$ as shown in Fig.~\ref{fig2}d. The comb $\mathcal{G}$ consists of parallel graphene stripes alternating
with metallic split-gate stripes G$_{\mathrm{A(B)}}$ which form the other
two combs. The G$_{\mathrm{A}}$ and G$_{\mathrm{B}}$ combs serve as
split-gates to creating of the transverse electric field $\mathbf{E}\neq 0$
across $\mathcal{G}$. Additionally, there are three other metallic combs, C$%
_{\mathrm{e}}$, C$_{\mathrm{h}}$ and H, which are oriented in perpendicular
to the former $\mathcal{G}$ and G$_{\mathrm{A,B}}$ combs. The two combs, C$_{%
\mathrm{e}}$ and C$_{\mathrm{h}}$, act as heat sinks, while the H comb
represents a heat source. Besides, there are two more back gate combs, G$_{%
\mathrm{L}}$ and G$_{\mathrm{R}}$, which are placed underneath of the
dielectric substrate and which are also oriented in-perpendicular to the G$_{%
\mathrm{A,B}}$ and $\mathcal{G}$ combs.

\begin{figure}
\includegraphics[width=85 mm]{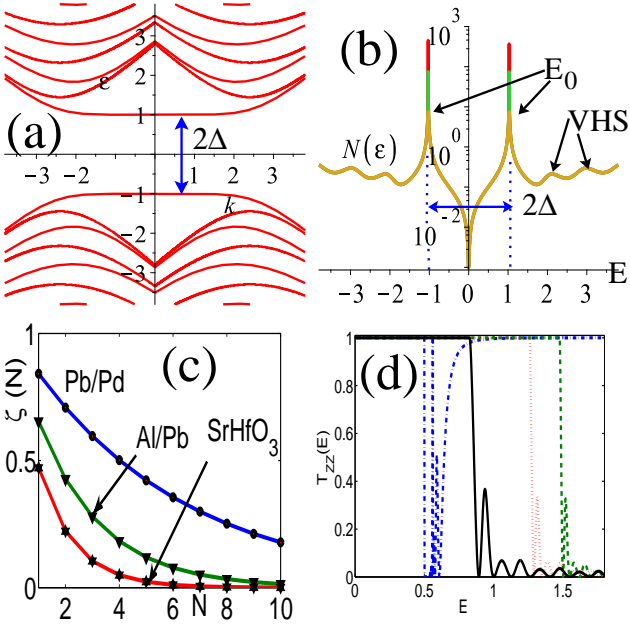}
\caption{ {Color online. (a)~The electron energy subbands in the graphene stripe ${\cal G}$ with zigzag edges. The transverse split gate voltage $V_{\rm SG}$ induces an energy gap $2\Delta= e V_{\rm SG} = e (V_{\rm GA} - V_{\rm GB})$ which splits the narrow zero-energy level. (b) Density of electron states $N_{\varepsilon }$ for three inelastic scattering rates $\gamma_{1,2,3} = 0.01$, $0.002$ and $0.0004$ in units of $\Delta $. The two sharp peaks constitute the Heavy Chiral Fermion (HCF) excitations arising at energies $\pm \Delta $. The corresponding peak height $N (\varepsilon = \pm \Delta)$ depends on $\gamma $.  It exceeds the  Van Hove singularity (VHS) peaks $N_{\rm VHS}$ by the big factors $N (\varepsilon = \pm \Delta)/N_{\rm VHS} = \sqrt{m^*/m_e} = 10$, and  100. (c)~The phonon transmission coefficient $\zeta $ for multilayered metallic electrode H (shown in Fig.~\ref{fig2}c) versus the number of layers $N$. (d)~The transmission coefficient $T_{zz}^{RR/{\cal G}}(\varepsilon )$ between the RR and uncovered ${\cal G}$ sections for different electron subbands.}}
\label{fig3}
\end{figure}

A temperature gradient through the H/RR-interface excites a heat flow carried by uncorrelated electrons and holes moving in the same direction. When the electron and hole concentrations are equal (as it happens in the graphene RR region) there is no electric current $I_{\rm th} = 0$ and no Seebeck bias voltage $V_{\rm th}$ is generated across the H/RR-interface. However there is a visible energy transfer from the ``hot" to the "cold" sides. 
The non-equilibrium electron and hole excitations in the RR-region result from the thermal injection of the electrons and holes and is caused by the temperature gradient across the H/RR-interface. 
The thermally injected electrons and holes cross the interface in the same direction and in equal number.  The electron and holes are non-correlated and do not form excitons, but the net electric charge which they transfer through the interface is equal to zero. In this way, the H/RR-interface lets no electric current through although the thermal injection of electrons and holes generates a sizable electric current in the ${\cal G}$-stripe. 
Then the H/RR-interface behaves as having no electric conductance. However if instead of the temperature gradient a finite bias voltage is applied, then a finite electric conductance emerges indeed.
Because the HCF-singularity is very sharp, and because there is an energy gap, the energy recombination time is very long which causes the electron excitations to be accumulated at the upper HCF level at $\varepsilon = \Delta$ while the hole excitations accumulate at $\varepsilon = -\Delta$. The electron and hole distribution functions are strongly non-equilibrium.

Both the electrons and holes are supplied into the electrically neutral RR by the thermal injection from H. Then the electric current $\mathbf{J}$ in FET$_{\rm R}$ actually emerges due to the HCF holes propagating from the RR region toward the C$_{h}$ electrode. Simultaneously, the electric current $\mathbf{J} $ in FET$_{\rm L}$ also consists of HCF electrons propagating in the opposite direction from RR toward C$_{e}$. In this way, both the HCF electrons and holes transfer the heat from the hot (H) to the cold (C$_{\mathrm{e,h}}$) electrodes. Then the combined thermal flow $Q_{\mathrm{th}}$ of electrons and holes generates an electric current accompanied by a finite Seebeck voltage drop $V_{\mathrm{SD}} = \sum_{n}V_{\mathrm{SD}}^{n}$. All the $\mathcal{G}$-stripes are connected in parallel, thus the net electric current through the whole TEG is $I = \sum_{k}I_{k}$ where $I_{k}$ is the current through the $k$-th $\mathcal{G}$-stripe. The Seebeck voltage drop on the two FET$_{\rm L,R}$ is $V_{\mathrm{SD}}^{n} = 2 {\cal S}_n \delta T_n $. Because the heat flow is transferred by heavy charged HCF particles ($m^*/m_e >> 1$) it ensures a considerable value of Seebeck coefficient ${\cal S}_n$ (see below). In turn it yields a high thermoelectric voltage $V_{\rm SD}^{n} = 2{\cal S}_{n} \delta T_{n}$ generated by the heat flow when a finite temperature difference $\delta T_{n}$ is maintained between the H and C$_{\rm e(h)}$ electrodes of the $n$-th $\cal G$-TEG. Then the $\cal G$-TEG generates the electric power $Q_{\rm el}^{{\cal G}-TEG} = \sum_{k,n}I_{k} V_{\rm SD}^{n}$ which is considerably higher than in conventional TEG devices. (See more details in \cite{Shaf-Arx},~A-I).

\begin{figure}
\includegraphics[width=85 mm]{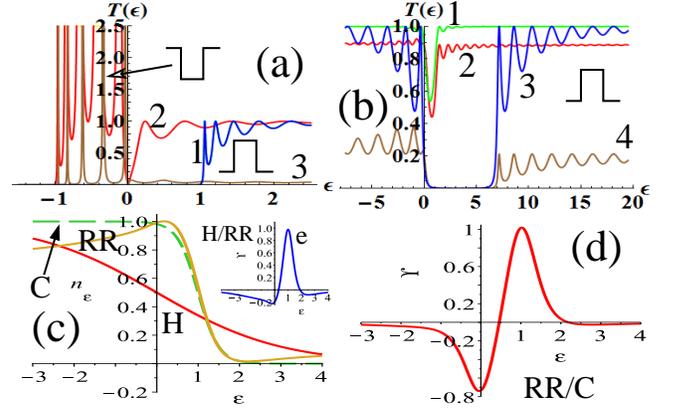}
\caption{ {Color online. (a) The transmission coefficient ${\cal T}({\varepsilon })$ for conventional electrons and conventinal heavy fermions through the well and potential barrier. (b) ${\cal T}({\varepsilon })$ for chiral fermions: curves 1, 2 correspond to regular chiral fermions while curves 3, 4 to the HCF excitations. Curves 1, 3 are for the incident angle $\varphi =\pi/8$ while curves 2, 4 are for $\varphi =3\pi/8$. (c)~The electron distribution functions in the cold electrode (C), recombination region (RR), and in the hot electrode (H). Inset shows the driving factor $\Upsilon_{\mathrm{H/RR}}^{(e)} (\varepsilon)$ for the tunneling current through the H/RR-contact. (d)~The driving factor $\Upsilon_{\mathrm{RR/C}}^{(e)} (\varepsilon)$ for electrons and holes propagating along the $\mathcal{G}$-stripe.}}
\label{fig4}
\end{figure}

Another increase of $Z \delta T$ is accomplished when the phonon part of the heat conductance $\Lambda _{\mathrm{ph}}$ is strongly reduced. A visible decimation of $\Lambda _{\mathrm{ph}}$ by a few orders of magnitude is achieved, e.g., if the hot H-electrodes are metallic multilayers fabricated as sketched in Figs.~\ref{fig2}c and Figs.~1-A a,b. The multilayered structure shown in Fig.~\ref{fig2}c and Figs.~1-A a,b serves for decimating of the phonon transport through the TEG by the factor $\zeta $ which we plot in Fig.~\ref{fig3}c versus the number of layers $N$. The transmission probability of a phonon through a single elemenetary ABA crank is presented in Fig.~1-A c (see \cite{Shaf-Arx}). From Fig.~1-A c one might notice that the electron transmission probability ${\cal T}_{\rm el}$ through the metallic multilayer remains almost ideal ${\cal T}_{\rm el} \simeq 1$ while the phonons strongly attenuate since $\zeta << 1$. Because the multilayered metal contacts H and C$_{\rm e,h}$ are reducing the phonon component of the heat flow, the figure of merit $Z \delta T$ can be increased additionally by orders of magnitude (see \cite{Shaf-Arx},~A-II).

The high thermal conductance $\Lambda_{\rm ph} $ due to propagating the phonons in graphene represents another obstacle to achieving a noticable TEG efficiency. One can strongly reduce $\Lambda_{\rm ph} $, e.g., by connecting the graphene stripe in a sequence with a material which  $\Lambda_{\rm ph} $ is very low whereas the electron part, $\Lambda_{\rm e} $, is high. 
Here we suggest to decimate the phonon part of heat conductance by connecting the graphene stripe in a sequence with a multilayered valve pad. A multilayer with random layer thickness had been considered by authors of Ref.~\cite{Musho} who studied a set of semiconducting layers with random thickness where energy gaps in the phonon spectrum are formed. However there are several basic complications when using the method \cite{Musho}. ({\it i}) The authors of Ref.~\cite{Musho} considered the multilayer as a set of semiconducting layers with random thickness where energy gaps in the phonon spectrum are formed. Since the phonon gaps in distinct layers are positioned at different phonon energies, the resulting phonon heat conductance is strongly suppressed. Although it is supposed to improve $Z T$, the idea \cite{Musho} actually fails to work properly. The controversy originates from two inherent contradictions. Namely, if the semiconducting layers are strongly coupled, the electric conductance and Seebeck coefficient are big. However, on the other hand, a strong interlayer coupling eliminates the phonon gaps since series of strongly coupled layers behave like a bulk material where the phonon gaps vanish. It causes the phonon heat conductance $\Lambda_{\rm ph}$ not to be small. In the opposite limit, when layers are weakly coupled, the phonon gaps are indeed formed, causing the phonon heat conductance being suppressed. But unfortunately the electric conductance and Seebeck coefficient are diminished with it also, because they are proportional to the interlayer coupling strength. In either case, weak or strong interlayer coupling, the figure of merit cannot be improved considerably by using the method of Ref.~\cite{Musho}. Our approach is based on a different idea which is more suitable for the TEG optimization. First, we consider {\it metallic} multilayer with ideally transparent interfaces. The metals have a much higher electron density of states (DOS) than the semiconductors do. Besides, the metal/metal interfaces are very transparent for electrons. Therefore the lateral electric conductance and Seebeck coefficient through the multilayer is much higher than of the semiconducting multilayer with decoupled layers as in Ref.~\cite{Musho}. Another critical distinction is that we use two metals A and B with similar Fermi velocities $v^{\rm A}_{\rm F} \simeq v^{\rm B}_{\rm F}$ but with very different sound velocities  $s_{\rm A} << s_{\rm B}$ (e.g., Pb/Al or Pb/Sn). It yields a strong reduction of the phonon thermal conductance $\Lambda _{\rm ph}$ without sacrificing the electron transport coefficients and thus it ensures a considerable improving the TEG figure of merit. Another benefit of using the metallic multilayer instead, e.g., the semiconducting multilayer \cite{Musho} is that the much higher electron density of states in metals much better matches the high density of states in the HCF peaks.
Because the electron part of the heat flow $Q_{\rm el}$ is considerably increased while the phonon part of the of the thermal flow $Q_{\rm th}$ is decimated, one might increase $Z T$ by orders of magnitude.  

One computes the net electric power generated by the whole TEG with equivalent ${\cal G}$-stripes and equivalent ${\cal G}$-TEGs as  $Q_{\rm electr} = N_{\cal G} \times I V_{\mathrm{SD}} = N_{\cal G} \times G_e (2 {\cal S} \delta T)^2$ where $N_{\mathcal{G}} $ is the total number of  $\mathcal{G}$-TEG elements in the TEG array and we have omitted index $n$. The corresponding thermal power is $Q_{\mathrm{heat}}= N_{\cal G} \times \Lambda  \delta T$. Both the quantities, $Q_{\rm electr}$ and $Q_{\mathrm{heat}}$, are expressed in terms of Seebeck coefficient ${\cal S} $, temperature differences $\delta T$, electric $G_e$, and thermal $\Lambda $ conductivities of the individual $\mathcal{G}$-TEG sections. The TEG conversion efficiency is then estimated as $\{Z \delta T \}_{\mathrm{TEG}}=Q_{\rm electr}/Q_{\mathrm{heat}} = 4 {\cal S}^2 G_e \delta T/ \Lambda $. The electron part, $\Lambda_e$, of the thermal conductance $\Lambda $ is obtained as $1/\Lambda _{\rm e} = 2 /\Lambda _{\rm c}^{\rm e} + 1/\Lambda^{\rm e,cont}_{\rm G}$ and is typically $10^{3}-10^{4}$ times smaller than the phonon part, which is $1/\Lambda _{\rm ph} = 2/ \Lambda^{\rm ph} _{\rm c} + 1/\Lambda^{\rm ph} _{\rm G} + 1/\Lambda _{\rm{SiO}_{2}}$.~\cite{Cahill} Here $\Lambda$'s indices e, ph, c, G, and SiO$_2$ relate it to the electrons, phonons, contacts, ${\cal G}$-stripe, and dielectric SiO$_2$ substrate respectively.
There are only two conducting HCF channels per each ${\cal G}$-stripe. For one conducting channel in a regular graphene stripe by width $W = 10$~nm, one gets Seebeck coefficient at most $S \sim 10^{-4}$V/K. For a graphene stripe with zigzag edges by the same width $W\simeq $10 nm and the split gate voltage $V_{\rm SG} = 0.1$~V, where the HCF resonances are invented, one improves it to ${\cal S}\sim 10^{-3} - 10^{-1}$~V/K (see Fig.~\ref{fig1}). In combination with the multilayered ``hot" electrode it yields as much as $Z \delta T \sim 10^2$.

The electric power generated by a single $\mathcal{G}$-TEG is evaluated as $%
Q_{\mathrm{el}}\simeq \kappa_{\rm HCF} G_{q}V_{\mathrm{SG}}^{2}\simeq 1$~mW,
where $\kappa_{\rm HCF} =\sqrt{m^{\ast }/m_{e}}(\delta T_{\mathrm{H/RR}}/\delta T_{\mathrm{RR/C}})$, $G_{q}=2e^{2}/h=7.75\times 10^{-5}$~$\Omega^{-1} $, and we have used $m^{\ast }/m_{e} = 10^2$ (which corresponds to $\gamma = 0.05$~meV), $\delta T_{\rm H/RR} = 300$~K, $\delta T_{\rm RR/C} = 30$~K, $V_{\mathrm{SG}}=0.1$~V. Let us assume that the length of a single $\mathcal{G}$-TEG element is $L_{G}\simeq 2.5$~$\mu $m, the $\mathcal{G}$-stripe width is $W_{G}=10$~nm, and the period in the y-direction is $W_{p}=40$~nm. Then the TEG device by size 1~cm$\times $1~cm contains $N_{G}=10^{9}$ $\mathcal{G}$-TEG elements which  might generate the electric power about $1$~MW. One can see that the ${\cal G}$-TEG efficiency largely depends on the maximum value of $V_{\rm SG}$ and on reducing of $\Lambda_{\rm ph}$.
For practical realizations, instead of d.c. gate voltages, one can operate the TEG at industrial a.c. frequency $f = $ 60 or 50~Hz. The industrial a.c. power period is 10$^{10}$ times longer than typical timescale of the excessive heat dissolution. An extra heat in TEG is dissolved ten billion times faster than industrial voltage period $\sim $ 0.02~s. That's because the heat dissolution occurs on a time scale of the diffusion time or the energy recombination time. The diffusion time for relatively "clean" graphene with the RR-region size $L_{\rm RR} \simeq 1$~$\mu$m is $\sim 10^{12}$~s and the energy recombination time in graphene also is $\sim 10^{-12}~s$. Therefore the industrial voltage frequency cannot impact the TEG performance in no way. It can be used for generating the a.c. voltage using a steady heat source and the a.c. split gate voltages. The required a.c. $V_{\rm G}$ and $V_{\rm SG}$ might be generated immediately during the heat to electricity conversion process. 

Many graphene stripes often represent a combination of different sections with zigzag (G$_{\rm ZZ}$) and armchair (G$_{\rm AC}$) edges. The electron states from adjacent G$_{\rm ZZ}$ and G$_{\rm AC}$ sections are hybridized with each other. Because the quantized levels in the  G$_{\rm ZZ}$ and G$_{\rm AC}$ sections are localized at different energies, the inter-sectional ZZ/AC-coupling leads to broadening of the quantized states. It resuls in degrading the Seebeck coefficient and electric conductance and it also reduces the figure of merit.

The obtained result suggests that the $\mathcal{G}$-TEG devices which exploit filtering of the electric/heat currents and of using the voltage-controlled spectral singularities are able to improving of the thermoelectric conversion efficiency and the generated output electric power by 2-3 orders of magnitude as compared to presently known devices.
There are several new results in the work. They are ({\it i}) The suggested HCF singularities are much stronger than other types of spectral singularities (e.g., VHS singularities), therefore the HCF states add orders of magnitude more into the figure of merit ZT and into the generated electric power. The HCF peak position is controlled by the local gate voltage. ({\it ii}) The multilayered metallic pad with random thickness of layers provides another way of a substantial increase the figure of merit. 

I wish to thank P.~Kim, V.~Chandrasekhar, M.~S.~Dresselhaus, and G.~Chen for fruitful discussions. This work had been supported by the AFOSR grant FA9550-11-1-0311.

\end{document}


\title{Supplement: Heat to Electricity Conversion by a Graphene Stripe with Heavy Chiral Fermions}
\author{S.~E.~Shafranjuk}
\affiliation{Department of Physics and Astronomy, Northwestern University, Evanston, IL 60208}

\date{\today }

\maketitle

\section{A-I. Non-Equilibrium Thermal Injection}
The heat energy flow $Q^{\mathrm{eh}}$ from the ``hot" electrode H with temperature $T_{\mathrm{H}}$   into a much colder heat sinks $C_{\rm e(h)}$ proceeds along the $\mathcal{G}$-stripe with temperature $T_{\mathcal{G}}$ ($T_{\mathrm{H}}>>T_{\mathcal{G}}$). The process happens in two stages (\textit{i}) the thermal injection H$\rightarrow $RR and (\textit{ii}) heat/current transfer along the ${\cal G}-stripe$ as illustrated by the energy diagrams at the top of Fig.~2b (main text). In the diagram we show the distribution function of ``hot" excessive quasiparticles $\delta n_{\mathrm{e,h}}$ in H where $\delta n_{\mathrm{e}}=[n(\varepsilon /T_{\mathrm{H}})-n(\varepsilon /T_{\mathcal{G}})]\theta (\varepsilon -\varepsilon _{\mathrm{F}})$ corresponds to the thermally excited excessive electrons while $\delta n_{\mathrm{h}%
}=-[n(\varepsilon /T_{\mathrm{H}})-n(\varepsilon /T_{\mathcal{G}})]\theta
(\varepsilon _{\mathrm{F}}-\varepsilon )$ describes the excessive hole
thermal excitations [here $n(x) =1/(e^{x}+1)$]. In stage (\textit{i})~the temperature difference $\delta T = T_{\mathrm{H}} - T_{\mathcal{G}}$ initializes the non-equilibrium "thermal injection`` of ``hot" electrons and holes from H into the two levels $\pm \Delta $ localized inside the recombination region RR.  As a result, the upper $+\Delta $ level becomes inversely populated by the non-equilibrium electrons while the lower level $-\Delta $ is populated with non-equilibrium holes. The process is described by the quantum kinetic equation \cite{Keldysh}
\begin{widetext}
\begin{eqnarray}
\frac{\partial n_{\varepsilon }}{\partial t} &=&{\cal L}_{\mathrm{H/RR}}\left\{
\varepsilon ,n_{\varepsilon }\right\} +{\cal L}_{\mathrm{ep}}\left\{ \varepsilon
,n_{\varepsilon }\right\} +{\cal L}_{\mathrm{eh}}\left\{ \varepsilon
,n_{\varepsilon }\right\} +{\cal L}_{\mathrm{RR/C}_{\mathrm{e,h}}}\left\{
\varepsilon ,n_{\varepsilon }\right\}.  
\end{eqnarray}%
In the quasistationary case, one sets $\partial n_{\varepsilon }/\partial t \equiv 0$ which gives
\begin{eqnarray}
\Gamma _{\mathrm{H/RR}}\left( n_{\varepsilon }^{\mathrm{H}}-n_{\varepsilon
}^{\mathrm{RR}}\right) +\Gamma _{\mathrm{RR/C}_{\mathrm{e,h}}}\left(
n_{\varepsilon }^{\mathrm{RR}}-n_{\varepsilon }^{\mathrm{e,h}}\right) -\frac{%
n_{\varepsilon }^{\mathrm{RR}}-n_{\varepsilon }^{\mathrm{F}}}{\tau _{\mathrm{%
eh}}}-\frac{n_{\varepsilon }^{\mathrm{RR}}-n_{\varepsilon }^{\mathrm{F}}}{%
\tau _{\mathrm{ep}}} = 0,
\end{eqnarray}%
\end{widetext}
where we have used ${\cal L}_{\mathrm{H/RR}}=\Gamma _{\mathrm{H/RR}}\left( n_{\varepsilon }^{ 
\mathrm{H}}-n_{\varepsilon }^{\mathrm{RR}}\right) $ for the non-equilibrium
 thermal injection, ${\cal L}_{\mathrm{ep}}=\left( n_{\varepsilon }^{\mathrm{RR}}-n_{\varepsilon }^{\mathrm{F}}\right) /\tau _{\mathrm{ep}}$ is
the electron-phonon collision term, 
${\cal L}_{\mathrm{eh}}=\left( n_{\varepsilon
}^{\mathrm{RR}}-n_{\varepsilon }^{\mathrm{F}}\right) /\tau _{\mathrm{eh}}$
describes the electron-hole recombination, \cite{Rana}
and ${\cal L}_{\mathrm{RR/C}_{\mathrm{e,h}}}=\Gamma _{\mathrm{RR/C}_{\mathrm{e,h}}}\left( n_{\varepsilon }^{\mathrm{RR%
}}-n_{\varepsilon }^{\mathrm{e,h}}\right) $ accounts for the HCF electron/hole
escapes from RR into directions of C$_{\mathrm{e,h}}$. The corresponding
electron distribution functions are approximated as $n_{\varepsilon }^{%
\mathrm{H}}=1/\left( e^{\varepsilon /T_{\mathrm{H}}}+1\right) $ in the H
electrode and $n_{\varepsilon }^{\mathrm{e,h}}=1/\left( e^{\left(
\varepsilon \pm U_0\right) /T_{\mathrm{c}}}+1\right) $ in the C$%
_{e,h}$ shoulders where $U_0=V_{\mathrm{G}}/\alpha $ is the back gate-induced  shift of the electron electrochemical potential, $\alpha $ is the back gate efficiency, $n_{\varepsilon }^{\mathrm{F}}$ is the Fermi function.  Besides, $T_{\mathrm{H}\left( \mathrm{c}\right) }$ are the effective
electron temperatures in the H and C$_{\mathrm{e,h}}$-adjacent regions. The non-equilibrium distribution function $n_{\varepsilon }^{\mathrm{RR}}$ in the RR region then is%
\begin{equation}
n_{\varepsilon }^{\mathrm{RR}}=\frac{\Gamma _{\mathrm{RR/C}_{\mathrm{e,h}%
}}n_{\varepsilon }^{\mathrm{e,h}}-\Gamma _{\mathrm{H/RR}}n_{\varepsilon }^{%
\mathrm{H}}-\Gamma _{\varepsilon }n_{\varepsilon }^{\mathrm{F}}}{\Gamma _{%
\mathrm{RR/C}_{\mathrm{e,h}}}-\Gamma _{\mathrm{H/RR}}-\Gamma _{\varepsilon }}
\end{equation}%
where the electron energy broadening is $\Gamma _{\varepsilon }=\hbar / \tau _{%
\mathrm{eh}}+ \hbar /\tau _{\mathrm{ep}}$. If using Pd as a metal electrode one typically gets [11] $\Gamma _{\mathrm{H/RR}}=5$~meV for a rough $\mathcal{G}$%
/Pd-interface while $\Gamma _{\mathrm{H/RR}}=60$~meV for a smooth 
$\mathcal{G}$/Pd interface, $\Gamma _{\mathrm{RR/C}_{\mathrm{e,h}}}=100$~meV, $\Gamma _{\varepsilon }=1.5$~meV. The temperatures are taken as $%
T_{H}=630$~K, $T_{RR}=330$~K, and $T_{\mathcal{G}}=300$~K, the level position $\Delta = 50$~meV, which corresponds to the split gate voltage $V_{\rm SG}=0.1$~V.
The above formulas allows computing the electric and thermal currents. Both types of the currents are inhomogeneous in vicinity of the H and C$_{\rm e,h}$ electrodes on the corresponding spatial lengths $L_{\rm el} \sim 10-100$~nm and  $L_{\rm th} \sim 100-250$~nm. [11]  The ``hot" conventional electrons coming from H in RR are converted inside RR into the "heavy" HCF excitations during the short time [11] $\tau _{c}\simeq \hbar /\Gamma $ (in Ref.~[11] $\tau _{c}\simeq 10^{-13}$ s, due to the energy level broadening because the tunneling coupling between H and $\mathcal{G}$. One finds $\Gamma = hv_{\rm H}{\cal T}_{\rm H/RR}/(4 d_{\rm H}) \simeq 5$~meV where ${\cal T}_{\rm H/RR}$ is the H/RR interface transparency, $v_{\rm H}$ is the Fermi velocity in H, $d_{\rm H}$ is the contact barrier thickness). (\textit{ii})~The second stage consists of the {\it chiral} transport, RR$\rightarrow \mathcal{G}$, which occurs on the longer timescale  $\tau_{\rm esc} \leq L/v \simeq 10^{-12} $~s. Then most of the HCF electrons and holes inside RR are captured by the adjacent FET$_{\mathrm{L,R}}$. Simultaneously, minor fractions of HCF electrons and holes annihilate with each other~[14] during the time $\tau _{\rm eh}\geq 10^{-12}$ s [typically $\tau _{\rm esc}^{-1} \sim (3-7)\cdot \tau _{eh}^{-1}$]. It sets a requirement to the spatial dimension of RR as $L_{0}=v\tau _{\mathrm{eh}}\leq 1$ $\mathrm{\mu }$m. Since the thermal injection is essentially a non-equilibrium process, the transforming of electron states between the stages (\textit{i}) and (\textit{ii}) is incoherent.

\begin{figure}
\includegraphics[width=75 mm]{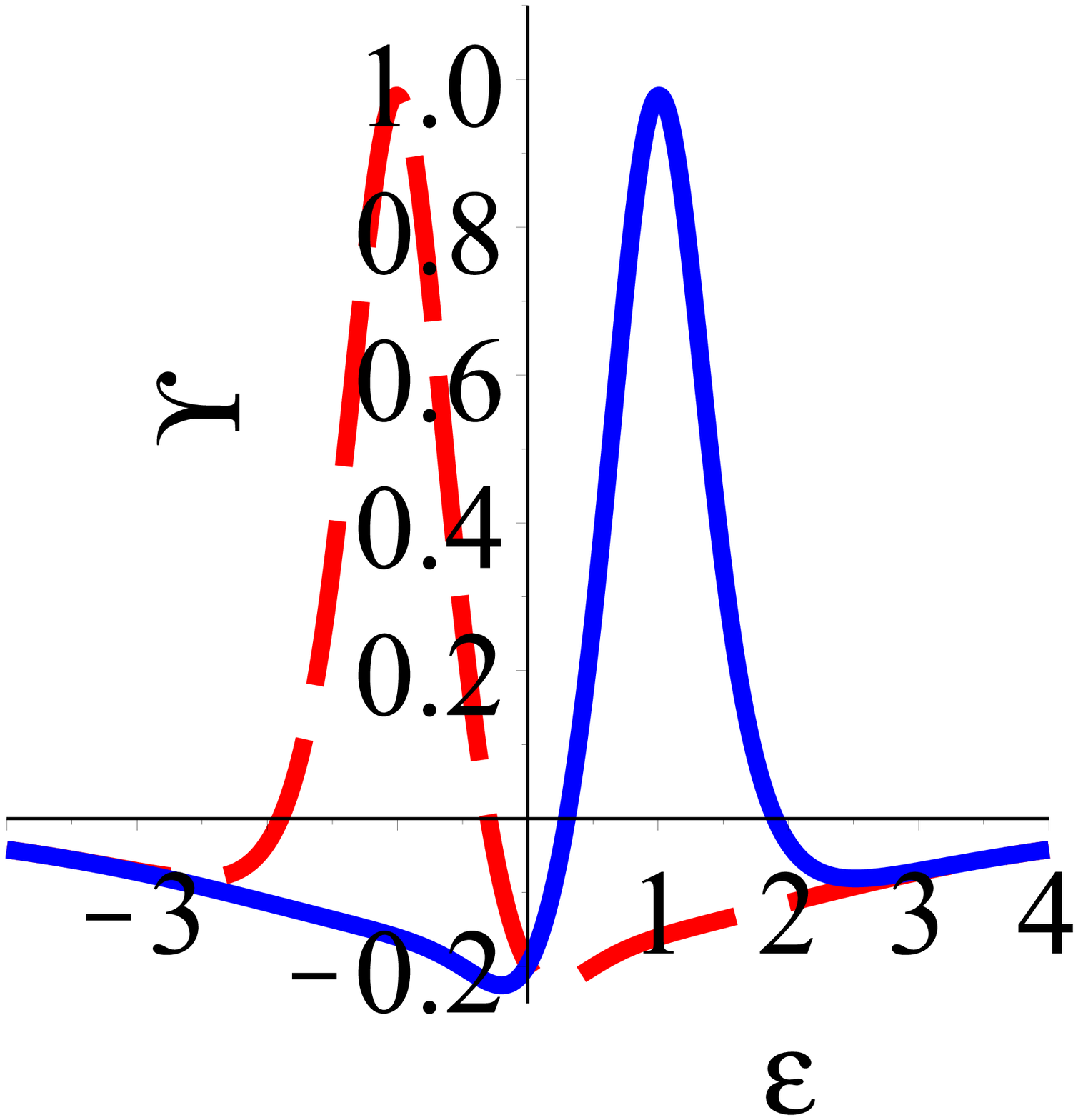}
\caption{{Color online.  The electron (solid blue curve) and hole (dash red curve) driving factors $\Upsilon_{\rm H/RR} (\varepsilon )$ where $\varepsilon $ is in units of $\Delta $.}} 
\label{fig2A}
\end{figure}

Knowing $n_{\varepsilon }^{\mathrm{RR}}$, one computes the electric conductance,  $G_e$, Seebeck coefficient, ${\cal S}$, and the electronic part of the heat conductance, $\Lambda _e$. The electric current through the H/RR-contact  vanishes (formally one gets $G_{\mathrm{H/RR}}\equiv 0$) because due to symmetry electron and hole excitations in graphene the electron part of the electric current is compensated by the hole part. For such reasons when computing the conductance for the whole TEG one formally sets the H/RR-contact conductance $G_{\mathrm{H/RR}}$ to zero. The physical meaning of this mathematical trick is that the RR-region is neutral (as the HH-region also is) because number of electrons there is equivalent to the number of holes. For the same reason there is also no finite bias voltage across the H/RR-contact ($V_{\rm H/RR} \equiv 0$). The H/RR {\it contact} Seebeck coefficient also vanishes since it is $\mathcal{S}_{\mathrm{H/RR}}= \mathcal{S}_{\mathrm{H/RR}}^{(e)}+\mathcal{S}_{\mathrm{H/RR}}^{(h)} \equiv 0$. Here $\mathcal{S}_{\mathrm{H/RR}}^{(e)} = -\left(
1/eT\right) \left( L^{\left( 1\right) }/L^{\left( 0\right) }\right) =V_{%
\mathrm{H/RR}}/\delta T_{\mathrm{H/RR}}$ where $V_{\mathrm{H/RR}}$ is the bias voltage (which is also equal to zero) and $\delta T_{\mathrm{H/RR}}$ is the temperature difference across the H/RR interface. Besides, due to the opposite electric charge of electrons and holes, one gets $\mathcal{S}_{\mathrm{H/RR}}^{(e)} = - \mathcal{S}_{\mathrm{H/RR}}^{(h)}$. Above we have introduced auxiliary functions $L^{\left( \alpha \right) }=\left(
2e^{2}/h\right) \int d\varepsilon \cdot \left( \varepsilon -\mu \right)
^{\alpha }M \left( \varepsilon \right) \mathcal{T} \left( \varepsilon
\right) \Upsilon_{\rm H/RR} (\varepsilon)$ where $ \Upsilon_{\rm H/RR} (\varepsilon) = -\partial \left[ n_{\varepsilon }^{\mathrm{H}}-n_{\varepsilon }^{\mathrm{RR}}\right] /\partial \varepsilon $ is  the driving factor, $\alpha = 1\ldots 3$, $M_{e}\left( E\right) = N_{G}\left( E\right) \left( \hbar v/L\right) $ is the number of modes, $N_{G}\left( E\right) $ is DOS shown in Fig.~3b (main text), $L$ is the $\mathcal{G}$-stripe length, $\mathcal{T} \left( \varepsilon \right) $ is the contact transparency. 
Unlikely to  $G_{\mathrm{H/RR}}$ and $\mathcal{S}_{\mathrm{H/RR}}$, which formally vanish at the H/RR contact, the contact thermal conductance $\Lambda _{%
\mathrm{H/RR}} = (L_{\mathrm{H/RR}}^{(2)}- [ L_{\mathrm{H/RR}}^{(1)}] ^{2}/L_{\mathrm{H/RR}}^{(0) })/(e^{2} T) = L_{\mathrm{H/RR}}^{(2)
}/(e^{2} T) $ remains essentially finite. A technical complication here is that the electron and hole excitation energies are represented by a non-analytical expression~(2). One can overcome that issue by using either of the tricks. The solution of Eq.~(2) near the HCF singularity is approximated by an analytical expression as we did at the end of Section~II.
Another trick is using the model form $M_{e}(\varepsilon ) = \sqrt{m^*/m_e} \cdot \delta ( \varepsilon -\Delta ) $ which gives $\Lambda _{\mathrm{H/RR}}\simeq \mathcal{F}_{\mathrm{H/RR}%
}\sqrt{m^*/m_e}( 2\Delta ^{2}/h)/ \delta T_{H/RR}$ where $\delta T_{H/RR} $ is an effective electron temperature difference across the $H/RR$-interface, and $%
\mathcal{F}_{\mathrm{H/RR}}=( -\partial[ n_{\varepsilon }^{%
\mathrm{H}}-n_{\varepsilon }^{\mathrm{RR}}] /\partial \varepsilon
) _{_{\varepsilon =\Delta }}\approx 0.1$.

The non-equilibrium injection is characterized by the electron driving factor $\Upsilon_{\rm H/RR} (\varepsilon)$ shown in inset Fig.~4c (main text). The holes are characterized by a very similar factor  but it depends on the energy differently: the minimum $\Upsilon_{\rm H/RR}^{\rm h} (\varepsilon)$ for holes occurs at the negative energy. We emphasize that both the driving factors, for electrons and holes, give pretty much the same $L^{(i)}$ for the electrons and holes. Therefore when computing $L^{(i)}$, one just have to use the electron/hole symmetry and remember that the electron and holes have opposite charge. This approach formally gives no electric current ($I=GV=0$) and no Seebeck effect ($V={\cal S}\delta T=0$) through the H/RR interface, while the heat flow due to tunneling electrons and holes is essentially finite. The situation is remarkably different for the electron and hole transport along the G-stripe. Since the electron and hole transports are separated from each other and proceed in opposite directions both, $I=GV \neq 0$ and $V={\cal S}\delta T \neq 0$, i.e., they are essentially finite. The electron/hole heat transfer is shunted by the phonon part of heat flow. However the latter is strongly diminished since the G-stripe is connected in sequence with the multilayer metallic pad with very low heat conductance  $\Lambda $.  
A very important consequence of the above properties is that the partial contact conductance $G$ and Seebeck coefficient ${\cal S}$ which are immediately related to the metal/graphene interface do not contribute into the $G$ and ${\cal S}$ entering $Z T$ for the whole TEG. That happens because the electric current flows only along the graphene stripes, and no electric current occurs between the ``hot" electrode and graphene in perpendicular direction. In contrast, the thermal flow occurs between the ``hot" metal electrode and graphene, then it splits into the two different directions. The electrons carry the heat to the source electrode while the holes transfer the heat to the drain electrode. The fraction of phonons which penetrate from the ``hot" electrode through the multilayered pad into the graphene stripe also splits into the two parts which carry the heat along the graphene stripe in the opposite directions.

There are at least two evident benefits of the tunneling injection from the ``hot" electrode into the HCF levels of the graphene stripe. ({\it a}) The sharp HCF singularities ensure a very good electric conductance and Seebeck coefficient along the graphene stripe. ({\it b}) From the thermal current stand point, the H/RR contact is connected in sequence with the graphene stripe. Next, the thermal conductance of the H/RR contact is much lower than the thermal conductance of the graphene stripe. Therefore the resulting net thermal conductance $\Lambda = (2\Lambda_{\rm H/RR}^{-1} + \Lambda_{\rm RR/c}^{-1})^{-1}$ is considerably diminished. Both the factors, ({\it a}) and ({\it b}), work toward considerable improving the TEG figure of merit.

Although the contact Seebeck coefficient ${\cal S}_{\rm H/RR}$ and the contact electric conductance $G_{\rm H/RR}$ through the H/RR interface formally vanish (i.e.,  ${\cal S}_{\rm H/RR} = 0$ and $G_{\rm H/RR} = 0$, as shown above), the contact thermal conductance $\Lambda_{\rm H/RR}$ remains essentially  finite.~[6,11] After being thermally injected from H into RR, the electrons and holes are quickly (during time $\sim 10^{-13}$ s) converted into the HCF excitations. In the RR region, the non-equilibrium HCF electrons and holes populate the levels $\pm \Delta $ inversely: the upper $+\Delta $ level is populated by excessive HCF electrons while the lower $-\Delta $ level by the excessive HCF holes.  

In the configuration shown in Fig.~2 (main text), the holes ballistically propagate from RR toward C$_{\mathrm{h}}$ while the electrons in the left $\mathcal{G}$-section proceed from RR toward C$_{\mathrm{e}}$. Thus, the latter stage implies  a chiral transmission of the excessive non-equilibrium HCF electrons from the upper $+\Delta $ level localized in RR into the upper $\Delta - U_{0}$ level located in the uncovered  $\mathcal{G}$-section adjacent to C$_{e}$ (see the diagram at the top of Fig.~2b, main text) where $U_0 \leq \gamma $, $\gamma $ is the localized level width. Simultaneously, the excessive holes are transmitted from the lower $-\Delta $ level in RR into the lower $-\Delta+U_{0}$ level localized near C$_{h}$. In this way, the full thermal flow $Q_{\rm el}$ from H to RR is eventually split between the C$_{h}$ and C$_{e}$ sections of the graphene stripe $\mathcal{G} $. In the uncovered ${\cal G}$-section, broadening $\gamma $ of the HCF level $E_0$ originates from coupling of the HCF states to the phonons. It yields $\gamma = \gamma_{\rm ph} = \hbar /\tau_{\rm ph}$ (typically $\tau_{\rm ph} \simeq 10^{-12}$~s at $T=10$~K). In the RR section, there is an additional coupling~[11] to the electron states in H which gives  $\gamma = \gamma_{\rm ph} + \gamma_{\rm H}$.

Along the ${\cal G}$-stripe, the TEG parameters $\mathcal{S}$, $G_{e}$, and $\Lambda _{e}$ are determined purely by the electron and hole transport. 
The underlying physical mechanism is the chiral
transmission of the HCF electrons and holes from the neutral RR section to the voltage p- and n-doped $\mathcal{G}$-sections. 
In the same approximation, one evaluates the electric conductance of FET$_{\rm L,R}$ along the ${\cal G}$-stripe between the RR and C$_{\mathrm{e,h}}$ as 
\begin{eqnarray}
G_{\mathrm{RR/C}} &=&L_{\mathrm{RR/C}}^{\left( 0\right)
}=\frac{2e^{2}}{h}\int d\varepsilon M\left( \varepsilon \right) \mathcal{T}_{%
\mathrm{RR/C}_{e(h)}}\left( \varepsilon \right) \Upsilon_{\rm RR/C} (\varepsilon)   \nonumber \\
&=&\frac{2e^{2}}{h}\sqrt{\frac{m^*}{m_e}}\mathcal{T}_{\mathrm{RR/C}_{e(h)}}\left(\varepsilon 
\right) \Upsilon_{\rm RR/C} (\varepsilon ) \vert_{\varepsilon = \Delta}
\end{eqnarray}%
where $\Upsilon_{\rm RR/C} (\varepsilon)$ is shown in Fig.~4d (main text). Analogously, one finds Seebeck coefficient along the ${\cal G}$-stripe 
$\mathcal{S}_{\mathrm{RR/C}}\simeq \left( \Delta
/e\right) /\delta T_{\mathrm{RR/C}} = V_{\rm SG} /\delta T_{\mathrm{RR/C}}$ and $\Lambda ^{\mathrm{RR/C%
}} =$ $L_{\mathrm{RR/C}}^{\left( 2\right) }/e^{2}T-TS_{e}^{2}G_{e}$ $\simeq 2\sqrt{m^*/m_e}$ $\left( \Delta ^{2}/\delta T_{\mathrm{RR/C}}\right)/h - 2\sqrt{m^*/m_e}$ $\left( \Delta ^{2}/\delta T_{\mathrm{RR/C}}\right)/h
\rightarrow 0$. The last result indicates that Seebeck coefficient $\mathcal{S}_{\mathrm{RR/C}}$ could be huge while the electron/hole part of the
thermal conductance $\Lambda_{\mathrm{RR/C}}$ along the stripe is typically  low. The phonon part
of the heat energy flow is $Q_{\mathrm{RR/C}}^{\mathrm{ph}}=\Lambda _{%
\mathrm{ph}}\delta T_{\mathrm{RR/C}}=N_{\mathrm{ph}}\kappa _{0}\delta T_{%
\mathrm{RR/C}}$ where $\kappa _{0}=5\times 10^{-12}$ W/K  at $T=10$ K
while the number of phonon modes $N_{\mathrm{ph}}$ also depends on the
temperature and the stripe geometry. 
The electron/hole heat energy flow is
\begin{eqnarray}
Q^{\mathrm{eh}} &=&\delta T^{2}\cdot S_{e\left( h\right) }^{2}G_{e\left(
h\right) }=\delta T^{2}\cdot S_{\mathrm{RR/C}}^{2}G_{\mathrm{RR/C}} 
\nonumber \\
&\simeq &\frac{2e^{2}}{h}V_{\mathrm{SG}}^{2}\cdot \left( \frac{\delta T_{%
\mathrm{H/RR}}}{\delta T_{\mathrm{RR/C}}}\right) ^{2}\sqrt{\frac{m^{\ast }}{%
m_{e}}} \nonumber \\
&=&\frac{2e^{2}}{h}V_{\mathrm{SG}}^{2}\cdot \kappa _{\mathrm{HCF}}
\end{eqnarray}%
where we have used $V_{\rm SG}= 2\Delta /e$, $\mathcal{S}_{\mathrm{RR/C}}=V_{\rm SG}/ \delta T_{%
\mathrm{RR/C}} $, $G_{\rm RR/C} = \sqrt{m^*/m_e} (2e^2/h) $, and we have defined the factor $\kappa_{\rm HCF}$. Because 
\begin{equation}
\kappa_{\rm HCF} = \left( \frac{\delta T_{%
\mathrm{H/RR}}}{\delta T_{\mathrm{RR/C}}}\right) ^{2}\sqrt{\frac{m^{\ast }}{%
m_{e}}}
\end{equation}
can be big, $\kappa_{\rm HCF} >> 1$, one might achieve huge values of $Q^{\mathrm{eh}}$.
Typically $\Lambda _{ph}^{\mathrm{H/RR}}<<\Lambda _{ph}^{\mathrm{RR/C}}$, therefore the $\mathcal{G}$-TEG net ``phonon" heat conductance~\cite{Cahill} is $\Lambda _{ph} \simeq \Lambda _{\mathrm{ph}}^{\mathrm{H/RR}}$,
which can be comparable to the contact electron heat conductance $\Lambda _{\mathrm{H/RR}} = 2\Delta ^{2} \sqrt{m^*/m_e} \mathcal{F}_{\mathrm{H/RR}} /(h\cdot \delta T_{\mathrm{H/RR}})$. 
It means that only the contact electron/hole and phonon heat conductances actually contribute into $\Lambda $. Summarizing the above estimates one arrives at $Z\delta T >> 1$.

\subsection{A-II. Blocking the phonon flow by multilayered electrodes.}
\subsubsection{Electric/heat conductance valve}
A general scenario for improvement the figure of merit $Z T$ implies
increasing the Seebeck coefficient $\mathcal{S}$ and electric conductance $G$
on one hand while reducing the heat conductance $\Lambda $ on the other
hand. A considerable raise of S and G is achieved by implementing the
quantized state resonances. The reducing of $\Lambda $ while preserving $%
\mathcal{S}$ and $G$ is accomplished when implementing of metallic
multilayers with the layer thickness randomly changed between $L\sim 10-100$ nm. The random layer thickness change is introduced to eliminate resonance transmission of phonons. Thus the phonon flow across the multilayer is decimated due to the low transmission probability $\zeta << 1$. However the contribution of phonons to the heat transfer might exceed the contribution due to the electrons and holes. The multilayer actually acts like a filter for the two components of microscopic heat transport.

The heat flow filtering mechanism is understood within a simple analytical
model. The multilayer represents a sequence of bilayers shown in Fig.~2a (main text) where metallic layers A and B are characterized by different Fermi ($v_{F}^{A,B}$) and sound ($s_{A,B}$) velocities. The phonon heat conductance through the multilayer $\Lambda_{\rm ph} $ is obtained from the Landauer formula. [27,28] One writes
\begin{eqnarray}
\Lambda _{\mathrm{ph}} &=& \frac{1}{2\pi \left( T_{\mathrm{H}}-T_{\mathrm{C}}\right) }\int_{0}^{\infty }\hbar \omega \zeta _{\omega }\left( N_{\omega }^{%
\mathrm{H}}-N_{\omega }^{\mathrm{C}}\right) d\omega
\label{Lph}
\end{eqnarray}%
where $T_{\mathrm{H,C}} $ and $N_{\omega }^{\mathrm{H,C}}$ are the temperatures and the Bose-Einstein distribution functions in the ``hot" and ``cold" ends. The phonon transmission probability $\zeta _{\omega }$ is obtained by the mode
matching method. \cite{Ando2} If the temperature gradient across the junction
is small, $T_{\mathrm{H}}-T_{\mathrm{C}}<<T_{\mathrm{H}}+T_{\mathrm{C}}$,
and the junction is ideally transparent for phonons, $\zeta \left( \omega
\right) =1$, $\Lambda _{\mathrm{ph}}$ is quantized \cite{Yama} as $\Lambda _{%
\mathrm{ph}}$ $\approx M\left( \pi ^{2}k_{\mathrm{B}}^{2}T/3h\right) $ where 
$M$ is the number of acoustic modes. For $L=1$~$\mu$m and $W = 20$~nm one roughly gets $\Lambda _{\mathrm{ph}} = \left( 50-6000\right)$~W/(m$\cdot $K).

Electron transport through the metallic multilayer is determined by the corresponding electrons ($\mathcal{T}_{el}$) and phonons ($\zeta $) transmission probabilities. We compute $%
\mathcal{T}_{el}$ and $\zeta $ in terms of the S-matrix method.~\cite{Datta} The electron S-matrix of the whole ABA$\dots $B multilayer is composed of elementary blocks ABA. Here we assume that there are no
interface A/B-barriers separating the A and B layers. The electron (hole) transmission ($t_{ABA}$) and reflection ($r_{ABA}$) coefficients which are the S$_{ABA}$ matrix elements are obtained from the A/B-interface boundary conditions as (see, e.g., Ref.~[4])%
\begin{eqnarray}
t\left( k_{A},k_{B}\right)  &=&2ik_{A}k_{B}/\mathcal{D}_{k_{A},k_{B}}  
\nonumber \\
r\left( k_{A},k_{B}\right)  &=&(k_{A}^{2}-k_{B}^{2})\sin (k_{B}L)/\mathcal{D}%
_{k_{A},k_{B}}  \label{tr}
\end{eqnarray}%
where the denominator is $\mathcal{D}_{k_{A},k_{B}}=\left(
k_{B}^{2}+k_{A}^{2}\right) \sin{(L k_{B})}+2ik_{A}k_{B}\cos{(L k_{B})}$, $k_{A}$ and $k_{B}$ are the electron wavevectors in A and B. Since the main contribution
to  $\mathcal{S}$ and $G$ comes from the electron states near the Fermi
level, one may use the linear dispersion law $E-E_{F}\simeq \hbar
v_{F}\left( k-k_{F}\right) $. Since the electron energy is conserved during
the interlayer transmission, $E^{A}=E^{B}$ it gives $\hbar v_{F}^{A}\left(
k^{A}-k_{F}\right) =\hbar v_{F}^{B}\left( k^{B}-k_{F}\right) $ or $%
v_{F}^{A}k^{A} \simeq v_{F}^{B}k^{B}$. This allows rewring of Eq. (\ref{tr}) simply
as%
\begin{eqnarray}
t_{a} &=&2i\alpha /\mathcal{D}_{\alpha }  \nonumber \\
r_{\alpha } &=&(\alpha ^{2}-1)\sin (\alpha )/\mathcal{D}_{\alpha }
\label{tr0}
\end{eqnarray}%
where now $\mathcal{D}_{\alpha }=\left( 1+\alpha ^{2}\right) \sin \alpha
+2i\alpha \cos \alpha $ and $\alpha =v_{F}^{A}/v_{F}^{B}$. The corresponding
phonon transmission and reflection coefficients are derived in a similar
way. 

\begin{figure}
\includegraphics[width=75 mm]{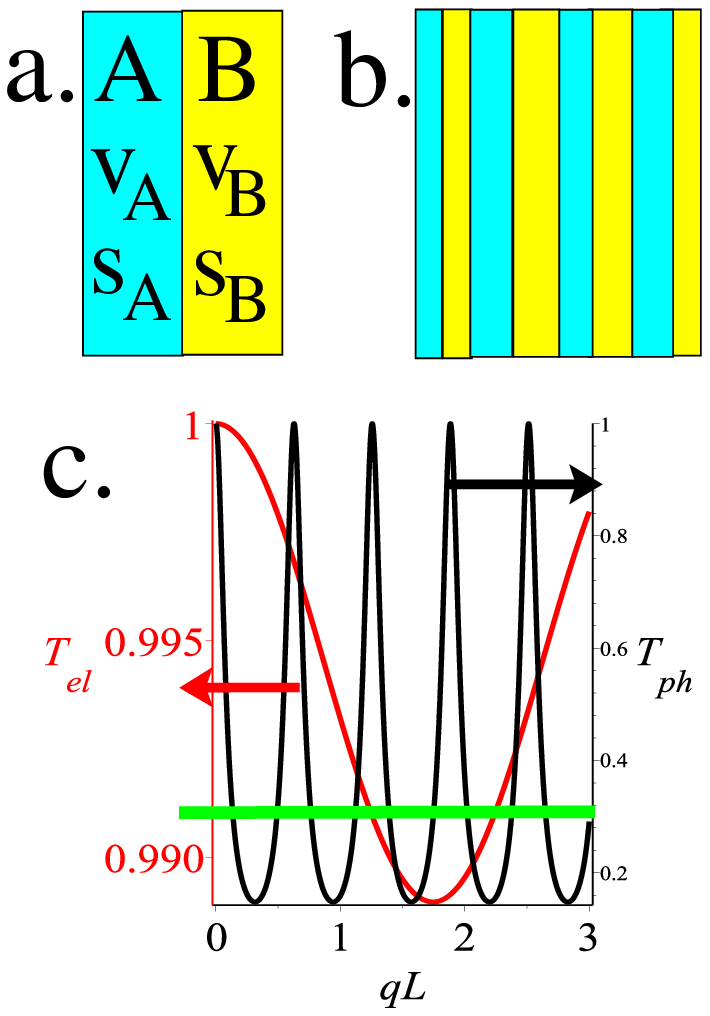}
\caption{{Color online. (a) The metallic bi-layer with $ v_{\rm F}^{\rm A}/v_{\rm F}^{\rm B} \simeq 1$ and $s_{\rm A}/s_{\rm B} >> 1$. (b) The metallic multilayer with randomized layer thickness. (c) The two-scale plot for a single elementary ABA crank with the layer thickness $L_{\rm A} = L_{\rm B} = 50$~nm and $\alpha = v_{\rm F}^{\rm A}/v_{\rm F}^{\rm B} = 0.9$ (for electrons, black curve) on one hand and for $\alpha = 5$ (for phonons, red curve) on the other hand. One may notice that phonons attenuate very strong while the electrons propagate almost free. The electrons correspond to $kL_{\rm A,B} << 1$ while phonons to $qL_{\rm A,B} \simeq 10$. For such reason the effective electron transmission probability is practically ideal, i.e., ${\cal T}_{el} \simeq 1$ while the averaged phonon transmission probability $%
\left\langle \zeta \right\rangle _{q} \simeq 0.3$. Further suppression of the
thermal phonon flow is accomplished by forming the sequence of $N$ bilayers
with a close but non-equivalent thickness around $L=50$~nm (for $T=10%
$~K). The phonon transmission through such a randomized
Pb/Al-multilayer is not phase-coherent, thus the total phonon transmission
probablity is composed of corresponding probabilities for
individual elementary bilayer blocks, i.e., $\zeta_{tot}\simeq
\zeta /N $ which already for, e.g., $N = 6$ gives $%
\zeta_{tot}\simeq 0.3/6=\allowbreak 0.05$ $\,$while $%
T_{el}^{tot}\simeq \left( T_{el}\right) ^{5}\simeq 1$, i.e., it remains
nearly ideal. The last exaple illustrates the principle of the heat flow
filtering: The electron and hole carry the heat free while the phonon
thermal conductance almost vanishes across the Pb/Al multilayer.}} 
\label{fig1A}
\end{figure}

A very interesting consequence of the above formula takes place when $\alpha
\simeq 1$ for electrons while simultaneously $\alpha >>1$ for phonons. Such situation takes place, e.g.,  when one of the metals in the elementary bilayer is lead (i.e., A=Pb) while another metal is aluminium (B=Al). The Fermi velocities for the lead and aluminium are very close ($v_{F}^{Pb} = 1.83\times 10^{6}$~m/s and $v_{F}^{Al} = 2.3\times 10^{6}$~m/s respectively) while the sound velocities in the two metals are quite different (i.e., $s_{Pb}=1158$~m/s and $s_{Al}=6420$~m/s). This gives a big difference between the two ratios $\alpha _{el} = v_{F}^{Al}/v_{F}^{Pb} = 0.8$ on one hand while $\alpha _{ph} = s_{Al}/s_{Pb} = 5.5$ on the other hand. The whole ABA$\dots $AB multilayer is then composed as a sequence of bilayers with randomly changing thickness. As is evident from Fig.~\ref{fig1A} c, the electron transmission probability through the whole metallic multilayer shown in Figs.~2c (main text), \ref{fig1A} a,b is nearly ideal, $T_{el}\simeq 1$ while the phonon transport is practically blocked since $T_{ph}<<1$.

Since the lattice constants of the two metals A and B differ, there is a lattice strain immediately at the layer's A/B-interfaces. However, the strained region involves just a few atomic layers at the Pb/Al or Pb/Sn interfaces. Therefore the strain affects just on a tiny fraction of the whole multilayer's volume. Contribution of that strained region into the electron and phonon transport across the metallic multilayer is negligible if the length of the strained region in lateral direction is much shorter than the metallic layer thickness or/and the electron and phonon mean free paths. Similar metallic multilayers had been widely used in the superconducting junction technology and it is well known the interface strains do not impact the lateral electron transport. That happens because it acts just like a scatterer which size is less than 1~nm, i.e., it is much narrower than the phonon's wave length of interest which exceeds 10 nm for $T >$ 300~K.
The strained region is too narrow to change the electron band structure on the local scale $< 1$~nm. The local short scale lattice deformation $< 1$~nm is also unable to generate an electrostatic potential barrier for the electrons propagating in the lateral direction.

The difference in transmission probabilities of the electrons and phonons
through the metallic multilayer can be exploited to filter the heat
transport components: The electrons and holes propagate through the metallic multilayer almost free while the phonon transport is considerably reduced.

The above simple analytical model is supported here by more rigorous numeric calculations. The filter pads are introduced to separate the ${\cal G}$-section both from the external electrodes and from the substrate thermally, but not electrically. We consider two different types of the heat/electric current valve pads. One design involves metallic multi-layers Pb/Al with the layer thickness $\sim 10 - 100$~nm. Another method is to depositing of pads made of SrHfO$_{3}$ and/or SrRuO$_{3}$. The layered materials have an appreciable electric conductance while their thermal conductance along the c-axis is remarkably low.~\cite{Keawprak,Maekawa} Planting of the H/RR pad between the metallic electrodes and graphene stripe would reduce the effective $\Lambda _{\mathrm{ph}}$ significantly, because the phonons which provide $\Lambda _{\mathrm{ph}}$ between the hot and cold ends are eliminated from the thermoelectric path. Then the net heat conductance which involves the path HOT $\Rightarrow$ RR $\Rightarrow$ ${\cal G}$ $\Rightarrow$ C$_{\rm e,h}$ $\Rightarrow$ COLD will be considerably diminished. Thus placing of the H/RR pad with a sufficient number of nanolayers allows to decimating of the phonon part $\Lambda _{\mathrm{ph}}$ the whole thermal conductance $\Lambda$.  
The optimal ${\cal G}$-TEG geometry is also determined by the electric and thermal transfer lengths which are estimated~[6,11] correspondingly as $L_{\rm el} \simeq 10-100$~nm and $L_{\rm th} \simeq 100-250$~nm.

\subsection{A-III. Numeric simulations.} The phonon part of the thermal transport through the TEG had been examined as follows. We describe the non-equilibrium heat flow through the ${\cal G}$-stripe in presence of multiple scattering on lattice defects, boundaries, and electrons. A finite temperature difference $\delta T$ between the opposite ends of each $%
\mathcal{G}$-stripe induces the thermal flow given as a sum of 
contributions of the individual phononic subbands.  The phonon density of states 
$F_{\beta} (\omega )$ related to the phonon subband $\beta$ is mismatched in adjacent layers of the H electrode sketched in Fig.~2c (main text).
Inside the ${\cal G}$-stripe, the phonon distribution function $N(\omega )$ is
non-equilibrium which means that  $N(\omega )$ deviates from the Bose-Einstein
distribution in the hot (H) and cold (C) ends. 
For a "clean" graphene stripe, the phonon mean free path exceeds the ${\cal G}$-stripe length $L$. Therefore the non-equilibrium effect does not influence the final results. 
The equilibrium phonon distribution at the  ${\cal G}$-stripe ends is established due to a free phonon diffusion into the bulk of attached metallic contacts and dielectric substrate. The thermal conductance $\Lambda_{\rm G}^{\rm ph}$ of the ${\cal G}$-stripe had been computed by using the phonon density of states $F_{\beta} (\omega)$ preliminary obtained for each phonon subband $\beta $.

The thermoelectric characteristics are found by solving the Dirac equation for chiral fermions in graphene (see above). The analytical model is verified by numeric calculations based on the density functional theory.~\cite{Fertig} The electron and phonon excitation spectra are obtained considering influence of the inelastic electron-phonon and elastic electron-impurity scatterings. 
They are taken into account along with
processes of the electron tunneling through the interface barriers. The electron-impurity and electron-phonon scatterings are included within the Keldysh-Green function technique~[25] which allows deriving of the quantum kinetic equations.

\subsection{A-IV. Transparency of the H/RR interface.} 
The thermal injection efficiency is directly related to transparency of the H/RR interface. The interface barriers which contribute into $\Lambda_{\rm H/RR}$ originate from the difference of the workfunctions in the metallic H electrode and the graphene $\mathcal{G}$-stripe right beneath of it [11,15]. The heat-conducting C$_{\mathrm{h}}$, H, and C$_{\mathrm{e}}$ electrodes are deposited at the top of the $\mathcal{G}$-stripe, as schematically shown in Fig.~2 (main text).
Another factor is change in the number of conducting channels
when electrons and holes tunnel from the 3D metallic H electrode into the 2D
graphene $\mathcal{G}$-stripe.~[21] Conversion of the
regular electrons and holes into the HCF excitations also contributes to $\Lambda_{\rm H/RR}$. Thus, for the $\mathcal{G}$-TEG, $%
\Lambda_{\rm H/RR}$ depends on the 3D/2D conversion efficiency $\eta $ and on the spatial distribution of charge carriers near the H/RR interface. The contact thermal conductance problem and its solution are illustrated in Fig.~4 (main text). In Fig.~4a (main text) we plot the transmission probability ${\cal T}\left( \varepsilon \right) $ as a function of the electron energy $\varepsilon $ for the conventional electrons penetrating a non-chiral potential barrier (curve 1), and the quantum well (curve 2).
Curve 3 shows ${\cal T}\left( \varepsilon \right) $ for the non-chiral heavy
fermions transmitting via a potential well. One can see that ${\cal T}(\varepsilon )$ is strongly suppressed in the latter case. For such a reason, the contact conductance for conventional "heavy`` fermions is low. Quite a different behavior ${\cal T}\left( \varepsilon \right) $ takes place if instead 
of the conventional "heavy`` electrons there are the  "heavy`` chiral (HCF) particles as is evident from Fig.~4b (main text). In Fig.~4b (main text) we compare ${\cal T}\left( \varepsilon \right) $ for the
conventional chiral fermions penetrating the chiral barrier (curves 1 and 3)
with the same characteristics for HCF particles (curves 2 and 4). One can
see that ${\cal T}\left( \varepsilon \right) $ is fairly good for both types of the chiral particles if the incidence angle is small, i.e., $\varphi =\pi /8$
(curves~1 and~2). For bigger incidence angles, i.e.,  $\varphi =3\pi /8$ (curves~3 and~4), for the HCF particles ${\cal T}\left( \varepsilon \right) $ becomes suppressed (curve~4). The electron thermal conductance $\Lambda_e $ of ${\cal G}$-TEG is determined by $2 \Lambda_{\rm H/RR}$. 
The dominant contribution into $\Lambda_{\rm H/RR}$ comes at the low angles $\varphi $, therefore using of the HCF particles helps to maintaining of  $\Lambda_e $ at a decent level.

\subsection{A-V. Graphene TEG parameters}
When evaluating the figure of merit and the electric power density of the
graphene TEG we admit the following parameters. The in-plane thermal
conductivity of graphene \cite{Balandin-1,Seol} is about $\Lambda_{\parallel } \sim$ 1000 W/(m$\cdot $K) and is determined by acoustic phonons. This is consistent with Refs.~[5,6], where for graphene layer thickness $h=0.4-4$ nm (which corresponds to $1<n<10$ layers), the in-plane thermal conductance is $\Lambda_{\parallel } = 1000$ MW/(m$^{2}\cdot $K).
Anisotropy of the thermal conductivity (in plane)/(out plane)  is about $\sim $1000 for the graphene/metal contact area which gives the cross-plane
thermal conductance $\Lambda_{\perp} \simeq 1$ MW/(m$^{2}\cdot$K). At $T=300$~K for Au/Ti/n-lG/SiO$_{2}$ multilayer one gets~[5,6] the thermal
conductance as $\Lambda_{\perp } = 20-30$ MW/(m$^{2}\cdot$K) while for Au/Ti/SiO$_{2}$ multilayer one obtains instead $\Lambda_{\perp } = 100$~MW/(m$^{2}\cdot$K). At the same time, the thermal
conductance for the 1-lG/SiO$_{2}$ interface (1-lG stands for the one-layer
graphene) is $\Lambda_{\perp } = 85$~MW/(m$^{2}\cdot$K). The electron heat conductance $\Lambda_{\rm el}$ has typically 3-4 orders of magnitude lower values than the phonon $\Lambda_{\rm ph}$ which gives $\Lambda_{el} << \Lambda_{ph}$. Therefore heat transfers along the graphene
plane and across the metal/graphene interface are carried predominantly by
phonons.

The in-plane sound velocity of graphene is $s  =1.5\times 10^{4}$~m/s. 
The Gaussian broadening $G(x,t_{2})=\exp (-x^{2}/t_{2}^{2})/(t_{2}\sqrt{\pi }%
)$ to model the random disorder potential in the channel is used with $%
t_{2}=\hbar v_{F}\sqrt{2\pi n_{0}}=117$~meV, corresponding to an
experimental $n_{0}$ (minimum sheet carrier concentration) of $n_{0}=5\times
10^{11}$~cm$^{-2}$.

There is a controversy in the literature concerning the electron-electron
collision time $\tau _{ee}$. Some authors regard $\tau _{ee}$ at room
temperatures as short as $\tau _{ee}=10^{-15}$ s which gives the mean free path $l_{ee}$ between the electron-electron scattering as $l_{ee}=v_{\mathrm{F}}\tau _{ee}=10^{-15}\times 10^{8}$~cm = 1~nm.
According to the experiment reported in Ref.~\cite{Bolotin}, an actual mean
electron free path in nanotubes at $T=5$~K is at least three orders of
magnitude longer and might far exceed 1 $\mu $m which gives the lower bond for $\tau _{ee}=l_{ee}/v_{\mathrm{F}}=10^{-12}$~s.

Broadening of the electron levels due to coupling of electron states in
adjacent regions we model here following Refs.~\cite{Perebei,Giovan,Nemec-1,Nemec-2}. According to ~\cite{Perebei,Giovan,Nemec-1,Nemec-2}, we use the broadening  caused by the $%
\mathcal{G}$/Pd-interface ramdomness as $\Gamma _{\mathrm{H/RR}}=5$~meV,
whereas $\Gamma _{\mathrm{H/RR}}=60$~meV for an ideal $\mathcal{G}$/Pd
interface, $\Gamma _{\mathrm{RR/C}_{\mathrm{e,h}}}=100$~meV, and $\Gamma
_{\varepsilon }=1.5$~meV. The temperatures are obtained self-consistently using the aforementioned values of the thermal conductance. Then we get $T_{H}=30$~meV and $T_{RR} = T_{\mathcal{G}}=5$~meV. In addition to the inhomogeneous broadening, graphene states under the metal have a homogeneous (lifetime) broadening $\eta $ due to coupling to the metal. Following to Ref.~\cite{Perebei}, we model the total broadening by a Voigt function  $\mathrm{Voigt}(x,t_{1},\eta )$, which is a convolution of a Gaussian of width $t_{1}$ and a Lorentzian of width $\eta $ (which is typically $\eta \simeq 5 \times 10^{-3}$~eV), and a cutoff energy $E_{0}$ of 1~eV, which is the typical bandwidth of d-bands in transition metals. The random potential disorder is expected to be much larger under the polycrystalline metal than in the channel. For a variation of the metal--graphene distance of 2~\AA\ (from 3~\AA\ to 5~\AA ), a 0.9~eV change in potential has been calculated.  

We also set the level position as $\Delta = 10 - 100$~meV which corresponds to the same value of the split gate voltage. The HCF peak width $\gamma $ is evaluated from know experimental data \cite{Bolotin} concerning the electron mean free path which in "clean" anealed graphene could be as much as $l_{\varepsilon } \sim 5$~$\mu$m which gives $\gamma = \hbar/\tau_{\varepsilon } =  \hbar v_{\rm F}/l_{\varepsilon } \leq 0.025$~meV. That corresponds to $\gamma / \Delta = 2.5 \times 10^{-4}$ and $m^*/m \sim 10^2$.

The quantum of thermal conductance at $T=300$~K is $\varkappa _{0} = \pi ^{2}k_{B}^{2}T/3h=1.5\times 10^{-10} \rm{W}/\rm{K}$ while at $T=10$~K it is $\varkappa _{0}=5\times 10^{-12}$~W/K.
The electric power generated by a single $\mathcal{G}$-TEG is evaluated as $Q^{\mathrm{eh}}\simeq \kappa _{g}G_{q}V_{\mathrm{SG}}^{2} = 7.75\times 10^{-4}$~W $\simeq 1$~mW where we used $V_{\mathrm{SG}}=0.1$~V, $G_{q}=2e^{2}/h=7.75\times 10^{-5}$~$\Omega^{-1}$ is the quantum conductance and we have introduced a large
dimensionless factor%
\begin{equation}
\kappa _{g}=\sqrt{\frac{m^{\ast }}{m_{e}}}\left( \frac{\delta T_{\mathrm{H/RR%
}}}{\delta T_{\mathrm{RR/C}}}\right) ^{2} \simeq 10^{3}
\end{equation}%
where we used the ratio of HCF and free electron mass as $%
m^{\ast }/m_{e}=100$ while the ratio of the temperature differences is obtained as $\delta T_{\mathrm{H/RR}}/\delta T_{\mathrm{RR/C}} = 10$.  The length $L_{G}$ of a single $\mathcal{G}$-TEG element is $L_{G}\simeq 2.5$~$\mu $m. The $\mathcal{G}$-stripe width is $W_{G}=10$~nm. The period of $\mathcal{G}$-stripe array in the y-direction is set as $W_{p}=40$~nm. Then the TEG device by size 1~cm$\times $1~cm contains $N_{G}=10^{9}$~$\mathcal{G}$-TEG elements which theoretically might generate the electric power 10$^{9}\times $ 1~mW = 1~MW.